\documentstyle[pre,aps,epsfig]{revtex}

\title{SECOND LAYER NUCLEATION\protect\\
       AND THE SHAPE OF WEDDING CAKES}

\author{J. Krug and P. Kuhn}
\address{Fachbereich Physik, Universit\"at Essen,
45117 Essen, Germany}

\begin{document}

\maketitle

\begin{abstract}
The rate of second layer nucleation -- the formation of a stable
nucleus on top of a two-dimensional island -- determines both the 
conditions for layer-by-layer growth, and the size of the top terrace
of multilayer mounds in three-dimensional homoepitaxial growth. It was
recently shown that conventional mean field nucleation theory overestimates
the rate of second layer nucleation by a factor that is proportional to 
the number of times a given site is visited by an adatom
during its residence time on the island.
In the presence of strong step edge barriers this factor can be large,
leading to a substantial error in previous attempts to experimentally
determine barrier energies from the onset of second layer nucleation.
In the first part of the paper simple analytic estimates of second layer
nucleation rates based on a comparison of the
relevant time scales will be reviewed. 
In the main part the theory of second layer
nucleation is applied to the growth of multilayer mounds in the
presence of strong but finite step edge barriers. 
The shape of the mounds is obtained by numerical integration
of the deterministic evolution of island boundaries, supplemented by a 
rule for nucleation in the top layer.
For thick films the shape
converges to a simple scaling solution. The scaling function
is parametrized by the coverage $\theta_c$ of the top layer,
and takes the form of an inverse error function cut off at
$\theta_c$. The surface width of a film of thickness $d$ is
$\sqrt{(1- \theta_c) d}$. Finally, we show that the scaling
solution can be derived also from a 
continuum growth equation. 

\end{abstract}

\section{Introduction}

An atom adsorbed on a crystal surface can move from one atomic
layer to another (usually from a higher layer to a lower one) only by
crossing a step. Irrespective of whether it occurs by simple hopping
or by concerted exchange, step crossing is often associated with an
additional energy barrier $\Delta E_{\mathrm S}$, which reduces the
effective rate of interlayer transport, $\nu'$, below the rate 
$\nu$ for adatom diffusion within one atomic layer 
\cite{Ehrlich66,Schwoebel66}. Here we are assuming
conventional Arrhenius laws for $\nu$ and $\nu'$,
\begin{equation}
\label{nu}
\nu = \nu_0 \; e^{-E_{\mathrm D}/k_{\mathrm B} T}, \;\;\;\;\;\;
\nu' = \nu_0' \; e^{-E_{\mathrm S}/k_{\mathrm B} T}
\end{equation}
and define $\Delta E_{\mathrm S} = E_{\mathrm S} - 
E_{\mathrm D}$.
The consequences for multilayer growth are
twofold. On the one hand, the confinement of atoms on top of the 
two-dimensional islands which form during the growth of the first layer
increases the nucleation rate of the second layer, which therefore may
take place before the first layer islands have coalesced, leading to
three-dimensional growth \cite{Tersoff94}. 
On the other hand, once three-dimensional mound-like features have 
formed through this (or some other) mechanism, the preferential attachment of atoms
to the ascending step of the vicinal terraces on the sides of the
mounds implies an effective uphill mass current, which stabilizes and
amplifies the mound formation \cite{Villain91}
(see Section \ref{Continuum}). The ubiquity of step
edge barriers on metal surfaces implies that also three-dimensional mound
growth should be ubiquitous, and indeed the phenomenon has been observed,
among other systems, in the homoepitaxy of Fe(100) \cite{Thuermer95,Stroscio95},
Rh(111) \cite{Tsui96}, Cu(100) \cite{Ernst94,Zuo97} and Pt(111) \cite{Kalff99}.

These patterns evoke questions about the kinetic processes that
determine the \emph{shape} of individual mounds \cite{Krug97,Politi97},
as well as about those processes that contribute to the 
\emph{coarsening} of the pattern by transferring mass from smaller mounds
to larger ones \cite{Tang98,Siegert98,Golubovic00,Michely01}. 
In the present contribution
we focus on the first set of questions. Starting from recent insights into
the process of nucleation on top of islands 
\cite{Krug00a,Heinrichs00,Krug00b,Castellano01}, we develop a minimalistic
model for the growth of a single mound, which allows us to analytically
determine the asymptotic mound shape, as well as to corroborate earlier
results that link the size of the topmost terrace of the mound to the
rate of interlayer transport \cite{Krug00a}. In the last section,
the asymptotic
mound shape is obtained as the solution of a continuum equation for the
growing surface. This extends a previously established connection between
macroscopic and atomistic levels of description \cite{Krug97} to a somewhat
more realistic situation.

\section{The Rate of Second Layer Nucleation}
\label{2nd}

\subsection{STRONG BARRIERS, IRREVERSIBLE AGGREGATION}

\begin{figure}
\label{Island}
\centerline{\epsfig{figure=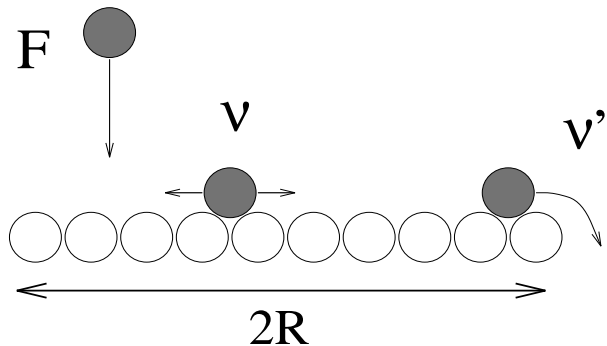,height=4.5cm,width=8.cm,angle=0}}
\caption{Kinetic processes involved in second layer nucleation. The figure
shows a circular island viewed from the side.}
\end{figure}

We consider the geometry illustrated in Figure 2. Atoms are
being deposited at rate $F$ onto a circular island of radius $R$. They
diffuse on the island with an in-layer hopping rate $\nu$, and
descend from it with an (average) interlayer hopping rate $\nu' < \nu$. 
We are looking for the probability per unit time, 
$\omega$, for a nucleation event to occur on the island. 
Assuming that dimers of adsorbed atoms are stable (as they will be
at sufficiently low temperatures), nucleation takes place as soon as
two adatoms meet. 

The kinetic rates $F$, $\nu$ and $\nu'$ combine to form three relevant
time scales\footnote{Throughout lengths will be measured
in units of the distance between adsorption sites on the surface.}: 
The \emph{interarrival time} 
\begin{equation}
\label{deltat}
\Delta t = \frac{1}{\pi R^2 F}
\end{equation}
between subsequent depositions onto the island; the \emph{diffusion time}
\begin{equation}
\label{tauD}
\tau_D \sim R^2/\nu
\end{equation}
required for an atom to diffuse once across the island, or for two atoms
to meet; and the \emph{residence time} $\tau$ which an atom spends on the
island before descending, when no nucleation event takes place. The calculation
of the residence time requires the solution of a stationary diffusion problem
with appropriate boundary conditions at the step edge, which yields 
\cite{Tersoff94,Krug00a}
\begin{equation}
\label{tau}
\tau = \frac{1}{2\nu} \left( R^2 + \frac{\nu R}{\nu'} \right).
\end{equation}
The residence time is comparable to the diffusion time $\tau_D$ if the 
suppression of interlayer transport is weak, in the sense that
$\nu/\nu' \ll R$, while in the 
opposite regime of \emph{strong step edge barriers},
$\nu/\nu' \gg R$, the residence time becomes $\tau = R/2 \nu'$ independent
of $\nu$. 

We will focus on the latter regime in the following. Then 
$\tau \gg \tau_D$, which implies that two atoms are certain to meet once
they are present simultaneously on the island. The probability for this to occur
is $\tau/(\tau + \Delta t) \approx \tau/\Delta t$ if $\Delta t \gg \tau$,
which is true for reasonable deposition fluxes. Multiplying this with the
total number of atoms deposited onto the island per unit time, we obtain
the expression \cite{Krug00a}
\begin{equation}
\label{omega}
\omega = \frac{\tau}{(\Delta t)^2} = \frac{\pi^2 F^2 R^5}{2 \nu'}
\end{equation}
for the nucleation rate, which is \emph{exact} under the stated conditions.

\subsection{COMPARISON WITH MEAN FIELD THEORY AND EXPERIMENTS} 

Equation (\ref{omega}) does not agree with the expression 
\begin{equation}
\label{omegamf}
\omega_{\mathrm mf} = \frac{\pi}{16} \frac{\sigma F^2 \nu R^4}{\nu'^2}
\end{equation}
derived from conventional (mean field) nucleation theory \cite{Tersoff94};
here $\sigma$ denotes a dimensionless capture number of order unity. 
In fact $\omega_{\mathrm{mf}}$ exceeds the true nucleation rate by a factor
which is of the order of $(\nu/\nu')/R \gg 1$. The reason can be traced
back to the fact that mean field nucleation theory treats the diffusing
atoms as noninteracting, which implies that they continue to diffuse 
even after having formed a dimer, thus accumulating too many nucleation 
events; in general the ratio
$\omega_{\mathrm{mf}}/\omega$ is proportional to the mean number of times
a site on the island is visited by an adatom during
its lifetime \cite{Castellano01}. 

In the analysis of experimental data, where
in the past the expression (\ref{omegamf}) has mostly been used instead
of (\ref{omega}), this overcounting implies that the value of the step
edge barrier $\Delta E_{\mathrm S}$ tends to be underestimated.
For example, for Pt/Pt(111) with CO-decorated steps, the value
$\Delta E_{\mathrm S} \approx 0.36 \; {\mathrm eV}$ was obtained using
(\ref{omega}), while the mean field expression (\ref{omegamf}) yields
only 0.28 ${\mathrm eV}$ \cite{Krug00a}. A reanalysis of a
second layer nucleation study on Ag(111) \cite{Bromann95} gave 
$\Delta E_{\mathrm S} \approx 0.15 \; {\mathrm eV}$ instead of the
estimate 0.12 ${\mathrm eV}$ reported in the original publication,
\emph{provided} the preexponential factor in (\ref{nu}) 
was assumed to have the conventional value
$\nu_0' = 10^{13} {\mathrm s}^{-1}$. The attempt to determine
both the step edge barrier
and the prefactor by fitting the Arrhenius form (\ref{nu}) to data
obtained at two different temperatures produces unphysically large
values for $\nu_0'$, which indicates that the data are not sufficiently
accurate for this purpose \cite{Krug01}. 
     
\subsection{REVERSIBLE AGGREGATION}

To some extent 
the above derivation
of the nucleation rate 
can be extended to the case of reversible aggregation,
where a minimum number of $i^\ast + 1 > 2$ atoms have to come together to form
a stable nucleus \cite{Heinrichs00,Krug00b}. The 
\emph{encounter time} required for the
atoms to meet is then no longer of the order of the diffusion time $\tau_D$,
but is (under the assumption that no metastable clusters exist) given by
\begin{equation}
\label{encounter}
\tau_{\mathrm enc} \sim R^{2 i^\ast}/\nu.
\end{equation}
For $i^\ast > 1$ it is therefore possible that $\tau_D \ll \tau \ll 
\tau_{\mathrm enc}$, which implies that nucleation is limited by the encounter
of the atoms, rather than by their simultaneous presence on the island.
The probability $p_{\mathrm enc}$ that the $i^\ast + 1$ atoms, once on
the island, meet before one of them escapes is then of the order
of $\tau/\tau_{\mathrm enc} \ll 1$, while $p_{\mathrm enc} \approx 
1$ when $\tau \gg \tau_{\mathrm enc}$. In general, the nucleation 
probability for a freshly deposited atom is proportional to the product
of $p_{\mathrm enc}$ and the probability that $i^\ast$ atoms are already
present on the island. Since the latter is of the order of 
$(\tau/\Delta t)^{i^\ast}$, the nucleation rate can be written as
\begin{equation}
\label{omegagen}
\omega \sim \frac{\tau^{i^\ast}}{(\Delta t)^{i^\ast + 1}} \; 
p_{\mathrm enc}.
\end{equation}
Interestingly, in the encounter limited regime where
$p_{\mathrm enc} \ll 1$, this is found to coincide in order of
magnitude with the prediction of mean field theory.
The evaluation of (\ref{omegagen}) in the two regimes defined
by the possible orderings of the relevant time scales  
($\tau_D \ll \tau_{\mathrm enc} \ll \tau \ll \Delta t$ or
$\tau_D \ll \tau \ll \tau_{\mathrm enc} \ll \Delta t$) 
then yields the expressions
\begin{equation}
\label{omega1}
\omega \approx \frac{1}{\Delta t}\left(\frac{\tau}{\Delta t}\right)^{i^\ast} 
\sim
F\left(\frac{F}{\nu'}\right)^{i^\ast} R^{3i^\ast + 2} 
\;\;\; {\mathrm for} \;\;\;\; R \ll 
\left(\frac{\nu}{\nu'}\right)^\delta
\end{equation}
\begin{equation}
\label{omega2}
\omega \approx \frac{1}{\tau_{\rm enc}}\left(\frac{\tau}{\Delta t}\right)^{i^\ast+1}
\sim
F\left(\frac{F}{\nu'}\right)^{i^\ast}\frac{\nu}{\nu'} \; R^{i^\ast + 3} 
\;\;\; {\mathrm for} \;\;\;
\left(\frac{\nu}{\nu'}\right)^\delta \ll R \ll \frac{\nu}{\nu'}
\end{equation}
where
\begin{equation}
\label{delta}
\delta = \frac{2}{2 i^\ast -1} < 1.
\end{equation}
Equations (\ref{omega1}) and (\ref{omega2}) respectively
describe the scaling regimes
II and III of \cite{Heinrichs00,Krug00b}.
In addition, a nonstationary regime I exists, where the residence time is effectively infinite and no stationary balance between deposition and escape of atoms from the island is established, as well as a regime
IV corresponding to weak step edge barriers.
These regimes play no role in the context
of the present paper.

\section{The Wedding Cake Model}

Our model for the shape of a single mound is a stack of concentric,
circular islands (Figure 3). The base is an island of
radius $R_0$ which does not grow, and the radius of the $n$'th island
is denoted by $R_n$.  

\begin{figure}
\label{Cake}
\centerline{\epsfig{figure=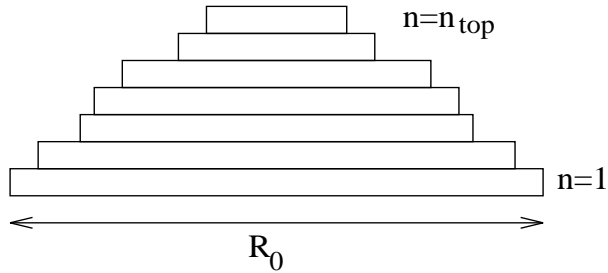,height=3.5cm,width=8.cm,angle=0}}
\caption{A stack of concentric circular islands seen from the side.}
\end{figure}

\subsection{RATE EQUATIONS FOR LAYER COVERAGES}

The rate at which the $n$'th island in the stack grows can be computed
from the solution of the stationary diffusion equation for the annular
region separating two islands \cite{Meyer95}. We simplify the problem even
further, and set the interlayer transport rate on the sides of the
mound to zero. This is motivated by the fact (to be demonstrated below)
that the mound steepens indefinitely, and hence the annular regions separating
two islands become very narrow. The probability for an atom deposited 
in such an annular region of width $l$ to attach to the ascending, rather
than the descending step is given by \cite{Krug97b}
\begin{equation}
\label{pplus}
p_+ \approx \frac{1/2 + \nu/\nu' l}{1 + \nu/\nu' l}
\end{equation}
which tends to unity for $l \ll \nu/\nu'$. Thus interlayer transport
will be increasingly suppressed as the mound steepens, and it seems
reasonable to neglect it from the outset\footnote{In addition,
setting $\nu' = 0$ makes it impossible for the boundaries of 
different islands in the stack to collide, which otherwise cannot
generally be ruled out \cite{Gossmann90}.}.

In the absence of interlayer transport the layer coverages 
$\theta_n = (R_n/R_0)^2$ satisfy the simple, \emph{linear} equations
\cite{Krug97,Cohen89}
\begin{equation}
\label{theta1}
\frac{d \theta_n}{dt} = F (\theta_{n-1} - \theta_n), 
\end{equation}
which express the fact that all atoms landing on top of the exposed
part of layer $n-1$ attach to the boundary of layer $n$. 
For the top layer $n_{\mathrm top}$ descent of atoms cannot be neglected -- 
since there are no sinks for atoms on the top layer, all atoms deposited
there have to attach to the descending step. The top terrace therefore
absorbs all atoms landing on layers $n_{\mathrm top}$ and 
$n_{\mathrm top} -1$, and grows according to 
\begin{equation}
\label{thetatop}
\frac{d \theta_{n_{\mathrm top}}}{dt} = F \theta_{n_{\mathrm top}-1}. 
\end{equation}

\subsection{NUCLEATION RULES}

Equations (\ref{theta1},\ref{thetatop}) have to be supplemented by a rule
for the nucleation of a new top terrace. A simple choice would be to posit
that nucleation occurs whenever the current top layer reaches some critical
coverage \cite{Meyer95}. 
More realistically, nucleation should be treated as
a stochastic process governed by the nucleation rate $\omega$ computed in 
Section \ref{2nd}. 
A deterministic rule which is close in spirit
to stochastic nucleation can 
be obtained as follows. Suppose the
current top terrace has nucleated at time $t=0$, and denote its radius
by $R_{\mathrm top}(t)$. Then the probability that no nucleation has
occurred on the top terrace up to time $t$ is given by \cite{Krug00a}
\begin{equation}
\label{P0}
P_0(t) = \exp \left(- \int_0^t dt' \; \omega(R_{\mathrm top}(t'))
\right),
\end{equation} 
and the probability density of the time $t$ of the next nucleation event
is $-dP_0/dt$. It follows that the mean value of $P_0$ at the time of 
nucleation satisfies 
\begin{equation}
\label{P0bar}
\bar P_0 = \int_0^\infty dt \; P_0(t) \left(- \frac{dP_0}{dt} \right) = 
1 - \bar P_0
\end{equation}
and thus $\bar P_0 = 1/2$. In the numerical implementation, we therefore
monitor the increase of 
$P_0$ during the growth of the top terrace and create a new
top terrace when $P_0 = 1/2$. For the nucleation rate 
$\omega$ the expression (\ref{omega}) will be used.

\subsection{APPROXIMATE ITERATIVE SOLUTION}

A full analytic solution of the model is difficult
because the nucleation probability (\ref{P0}) depends on the
size of the top terrace, which is determined by the entire
history of the wedding cake through the coupled equations
(\ref{theta1}). Some first insight 
can be gained by assuming that
the former top terrace ceases entirely to grow once
a new island has nucleated on top of it. This should produce
a lower bound on the island radii reached at a given time.

Denote by $R_{n-1}^{\mathrm top}$ the radius of island $n-1$ at
nucleation of island $n$, and by $t_{n}$ the time of
this nucleation event. Then during its tenure as the top
terrace, the radius of island $n$ grows, according to 
(\ref{thetatop}), as 
\begin{equation}
\label{Rntop1}
R_n(t) = \sqrt{F(t-t_{n})} R_{n-1}^{\mathrm top},
\end{equation}
and the time $t_{n+1}$ of the next nucleation event is 
determined by
\begin{equation}
P_0(t_{n+1}) = 
\exp \left( - 
\frac{\pi^2 F^2 (R_{n-1}^{\mathrm top})^5}{2 \nu'}             \int_{t_n}^{t_{n+1}} dt \; [F(t - t_n)]^{5/2}
\right) = 1/2.
\end{equation}
This implies a recursion relation
\begin{equation}
\label{Rntop}
R_{n}^{\mathrm top} = R_c^{5/7} (R_{n-1}^{\mathrm top})^{2/7}
\end{equation}
for the size of the top terrace at nucleation of the next layer,
where we have introduced the characteristic radius
\begin{equation}
\label{Rc}
R_c = \left( \frac{7 \ln 2}{\pi^2} \right)^{1/5} 
\left( \frac{\nu'}{F} \right)^{1/5}
\approx 0.868 \cdot \left( \frac{\nu'}{F} \right)^{1/5}.
\end{equation}
The time interval $\tau_n = t_{n+1} - t_n$ during which
$n_{\mathrm top} = n$ is related to  $R_{n}^{\mathrm top}$
through (\ref{Rntop1}), and correspondingly satisfies
\begin{equation}
\label{taun}
\tau_{n+1} = F^{-1} (F \tau_n)^{4/7}.
\end{equation}

It is easy to check that the recursion relations (\ref{Rntop})
and (\ref{taun}) approach fixed point values
\begin{equation}
\label{FP}
R_{n}^{\mathrm top} \to R_c, \;\;\;\;
F \tau_n \to 1
\end{equation}
exponentially fast in $n$. Thus asymptotically there is one nucleation
event during the growth of one monolayer\footnote{This has also been observed
experimentally for engineered mounds grown on 
Cu(100) \cite{Gerlach01}.}, and nucleation occurs when
the radius of the top terrace has reached the value $R_c$. 
We shall see below that these statements remain valid for the full
dynamics, although the approach of $R_{n}^{\mathrm top}$ and
$\tau_n$ to their asymptotic values is slower than exponential due
to the coupling to the lower layers.  

\subsection{NUMERICAL INTEGRATION}

Since the deposition flux $F$ and the size $R_0$ 
of the base of the mound
can be eliminated by rescaling the time and the island radii, 
the model effectively contains only the single
parameter $\nu'/F$. For convenience we will express this parameter
through the coverage $\theta_c$ of the top layer at nucleation,
which is given by 
\begin{equation}
\label{thetac}
\theta_c = \left(\frac{R_c}{R_0} \right)^2 = 
\left( \frac{7 \ln 2}{\pi^2} \right)^{2/5} 
\frac{(\nu')^{2/5}}{F^{2/5} R_0^2}.
\end{equation}

\begin{figure}
\label{CakeEvol}
\centerline{\epsfig{figure=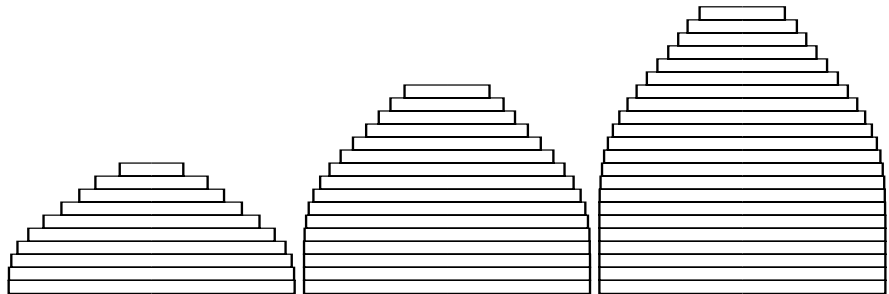,height=12cm,width=4cm,angle=-90}}
\caption{Wedding cakes grown numerically 
with $\theta_c = 0.1$, at total coverages of
$Ft = 5$, 10 and 15 monolayers.
The figure shows a stack of circular
islands seen from the side, i.e. the bar at level $n$ has a length
proportional to $\sqrt{\theta_n}$.}
\end{figure}

Figure 4 shows the result of a numerical integration of 
the model equations for $\theta_c = 0.1$. The figure suggests
that the number of exposed layers (with coverages $0 < \theta_n
< 1$) increases with time. This is quantified in Fig.5,
which shows the evolution of the surface width
\begin{equation}
\label{W}
W^2(t) = \sum_{n=1}^\infty (n - Ft)^2 (\theta_{n-1} - \theta_n).
\end{equation}
In the case of purely statistical growth with $\nu' = \theta_c = 0$,
$W^2 = Ft$ because the exposed coverages $\theta_{n-1} - \theta_n$
are given by a Poisson distribution with parameter $Ft$ 
\cite{Krug97,Cohen89}. Figure 5 shows that also for
$\theta_c > 0$ the squared width grows linearly in $Ft$, but with 
a coefficient that is less than unity;
in Section \ref{Asympt} it will be shown that the coefficient
is simply $1 - \theta_c$. 
Since the lateral mound size is fixed, the unbounded increase
of $W(t)$ implies that the mounds steepen indefinitely,
with the terrace width on the mound slopes decreasing as
$1/\sqrt{Ft}$; a quantitative expression is given in (\ref{lmax}).

\begin{figure}
\label{Width}
\centerline{\epsfig{figure=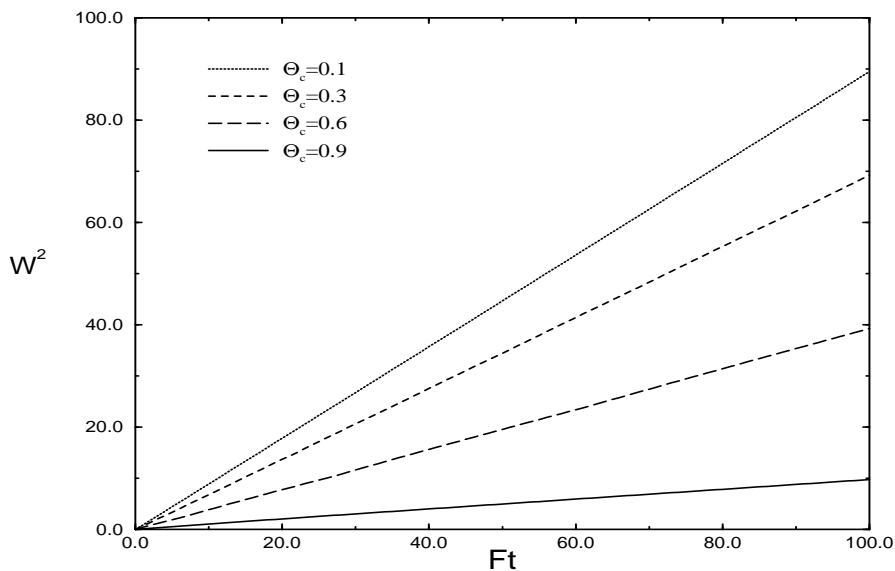,height=11.5cm,width=8cm,angle=-90}}
\caption{Surface width (\ref{W}) as a function of time.}
\end{figure}

Finally, Fig.6 illustrates the convergence of the top terrace
size at nucleation, and the time intervals between subsequent 
nucleation events, to their limiting values predicted by 
(\ref{FP}). As might be expected from the behavior of the
surface width, the convergence is linear in $1/\sqrt{Ft}$, rather
than exponential.  

\begin{figure}
\label{Converge}
\centerline{\epsfig{figure=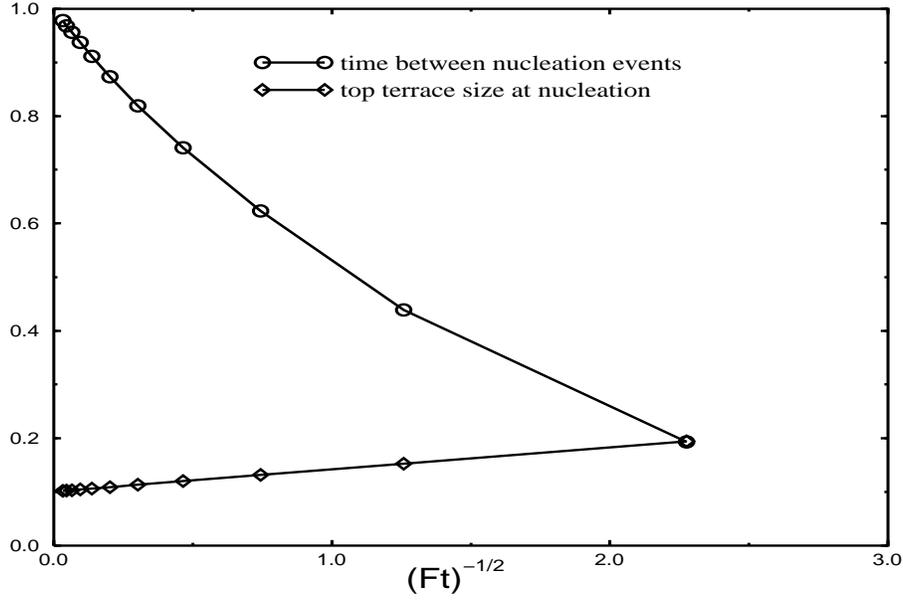,height=12cm,width=8cm,angle=-90}}
\caption{Convergence of the time between nucleation events, and
the top terrace size at nucleation, as a function of $1/\sqrt{Ft}$.}
\end{figure}

\subsection{THE ASYMPTOTIC MOUND SHAPE}
\label{Asympt}

\begin{figure}
\label{Shape}
\centerline{\epsfig{figure=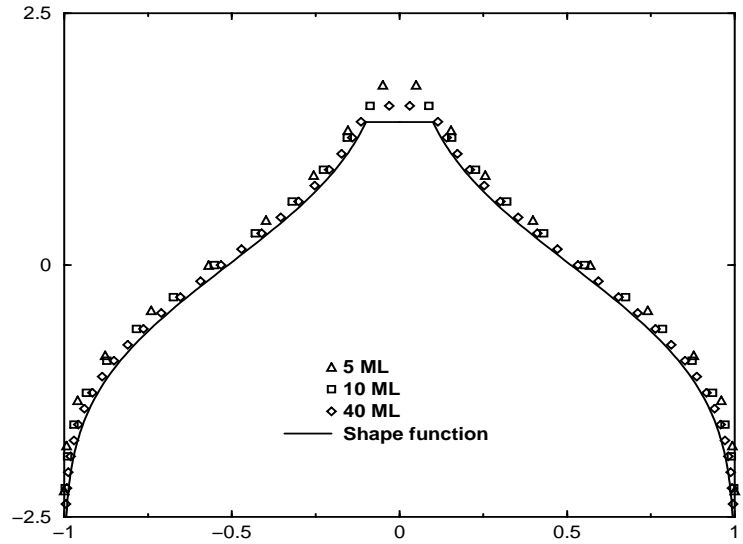,height=12.cm,width=9.cm,angle=-90}}
\caption{Convergence of numerically generated wedding cakes to the
asymptotic shape function $\Phi(x)$. The symbols show the coverage
profile rescaled according to (\ref{travel}).}
\end{figure}

The numerical solution of the model equations shows that
the height profile of the mound converges to a time-independent
asymptotic shape, when viewed relative to the mean film height
$Ft$ and rescaled horizontally by $\sqrt{Ft}$ to compensate
the increase of the surface width (Figure 7). To derive the shape 
analytically, we write 
the coverage profile $\theta_n(t)$ as a traveling wave
of width $\sqrt{Ft}$,
\begin{equation}
\label{travel}
\theta_n(t) = \Phi[(n-t)/\sqrt{Ft}].
\end{equation}
Inserting this into (\ref{theta1}) and expanding for large
$Ft$ one finds that the shape function $\Phi(x)$ has to satisfy
the differential equation
\begin{equation}
\label{Phi}
\Phi''(x) = - x \Phi'(x),
\end{equation}
where $x = (n-t)/\sqrt{Ft}$ is the scaling variable. 
This shows that the inflection point of the profile, where
$\Phi'' = 0$, is always located at $x = 0$, i.e. at 
$n = Ft$. The solution of (\ref{Phi}) which satisfies
the boundary condition $\lim_{x \to -\infty} \Phi(x) = 1$
reads
\begin{equation}
\label{error}
\Phi(x) = 1 - \sqrt{\frac{2}{\pi}}
C \int_{-\infty}^x dy \; e^{-y^2/2} = 
1 - C [1 + {\mathrm erf}(x/\sqrt{2})],
\end{equation}
where ${\mathrm erf}(s)$ denotes the error function, and $C$
is a constant of integration. The profile is cut off at the
rescaled height $x_{\mathrm max}$ of the top terrace, where
the coverage takes the value $\theta_c$, 
\begin{equation}
\label{xmax}
\Phi(x_{\mathrm max}) = \theta_c.
\end{equation}
Accordingly, the height of the top
terrace above the mean film thickness $Ft$ is 
$x_{\mathrm max} \sqrt{Ft}$. 

\begin{figure}
\label{fig_error}
\centerline{\epsfig{figure=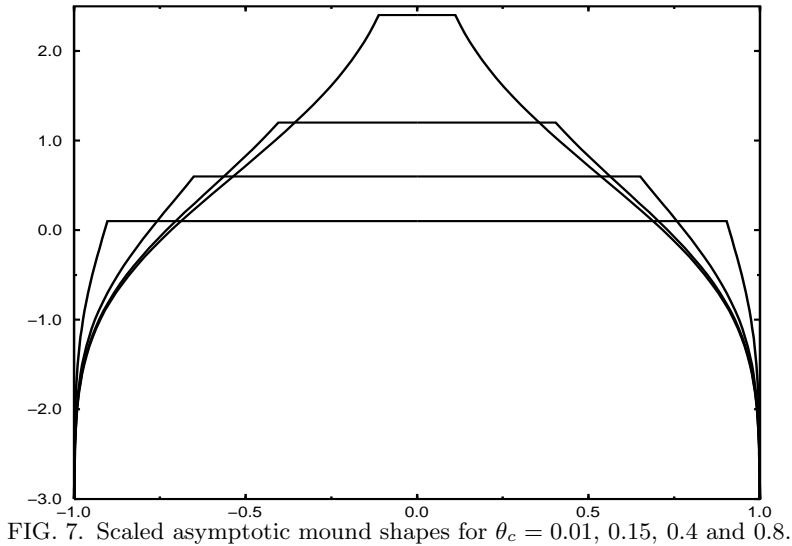,height=10.cm,width=7.cm,angle=-90}}
\caption{Scaled asymptotic mound shapes for $\theta_c = 0.01$, 0.15, 0.4
and 0.8.}
\end{figure}

\begin{figure}
\label{fig_parameters}
\centerline{\epsfig{figure=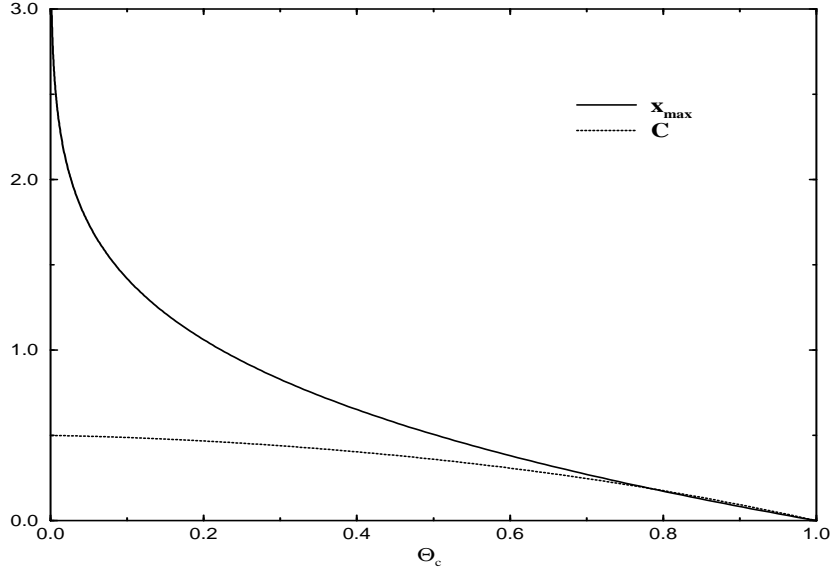,height=12cm,width=8cm,angle=-90}}
\caption{Shape parameters $C$ and $x_{\mathrm max}$ as a function
of $\theta_c$.}
\end{figure}

The two parameters $C$ and $x_{\mathrm max}$ of the shape function
are related by (\ref{xmax}), but to fix both of them a further
relation is required. This is provided by the normalization 
condition
\begin{equation}
\label{norm}
\sum_{n=1}^\infty \theta_n = F t,
\end{equation}
which, using (\ref{travel}), translates into
\begin{equation}
\label{phinorm}
\int_0^{x_{\mathrm max}} dx \; \Phi(x) = \int_{-\infty}^0 
dx \; (1 - \Phi(x)).
\end{equation}
Together equations (\ref{error},\ref{xmax},\ref{phinorm})
define a family of shape functions parametrized by
$\theta_c$. Examples of these functions are shown in 
Fig.8, while Fig.9 
illustrates the dependence of $x_{\mathrm max}$
and $C$ on $\theta_c$.
For the case of Pt  growth on Pt(111) at 440 K, the experimentally
determined mean mound shape has been compared to the scaling function
$\Phi(x)$ for $\theta_c = 0$, and good agreement was found except near
the top of the mound \cite{Kalff99}. Evaluation of the top terrace
size and comparison with (\ref{Rc}) yields the estimate
$\Delta E_{\mathrm S} \approx 0.25$ eV \cite{Michely99}. 

Other features of the mound shape can be easily extracted.
The surface width (\ref{W}) turns out to be given by
\begin{equation}
\label{W2}
W^2 \approx (1 - \theta_c) Ft
\end{equation}
asymptotically.  
The mound radius at the level of the mean film
thickness, $x = 0$, is given by $\sqrt{1 - C} R_0$, and 
the maximum terrace width on the mound slopes, located also
at the inflection point $x=0$, decreases with increasing film
thickness according to 
\begin{equation}
\label{lmax}
l_{\mathrm max}/R_0 \approx  
\frac{\Phi'(0)}{ 2 \sqrt{\Phi(0)}} \frac{1}{\sqrt{Ft}} = 
\frac{C}{\sqrt{2 \pi(1 - C)}} \frac{1}{\sqrt{Ft}} .
\end{equation}

\subsection{THE DISTRIBUTION OF TOP TERRACE SIZES}
\label{TopDist}

In an image of a multilayer film subject to the mound instability
one will typically see top terraces of a range of sizes,
because they are caught in different stages of their evolution.
In \cite{Krug00a} a simple model for the size distribution was
developed, which decouples the dynamics of the top terrace from the
rest of the mound by assuming that each top terrace grows on 
a template -- the base terrace -- of constant radius $R_b$. 
The nucleation probability (\ref{P0}) is then evaluated
with 
\begin{equation}
\label{Rtop}
R_{\mathrm top}(t) = \sqrt{Ft} \; R_b,
\end{equation} 
and its 
derivative $-dP_0/dt$ is interpreted as the probability distribution
for the time intervals between two subsequent nucleation events.
The base terrace radius is fixed through the requirement that the
mean time between nucleation events equals the monolayer deposition time,
which implies 
\begin{equation}
\label{mean}
F \int_0^\infty dt \; P_0(t) = 1.
\end{equation}
One obtains $R_b \approx 0.867 \cdot (\nu'/F)^{1/5}$, which is 
virtually indistinguishable from the expression (\ref{Rc}) 
derived from the condition $P_0 = 1/2$.
If one demands instead that nucleation takes place at the
maximum of the probability density, i.e. at the time determined
by $d^2 P_0/dt^2 = 0$, a prefactor of 0.873 is obtained.
All reasonable criteria give the same result, because
the probability density $-dP_0/dt$ is sharply peaked around
its maximum.

\begin{figure}
\label{SizeDist}
\centerline{\epsfig{figure=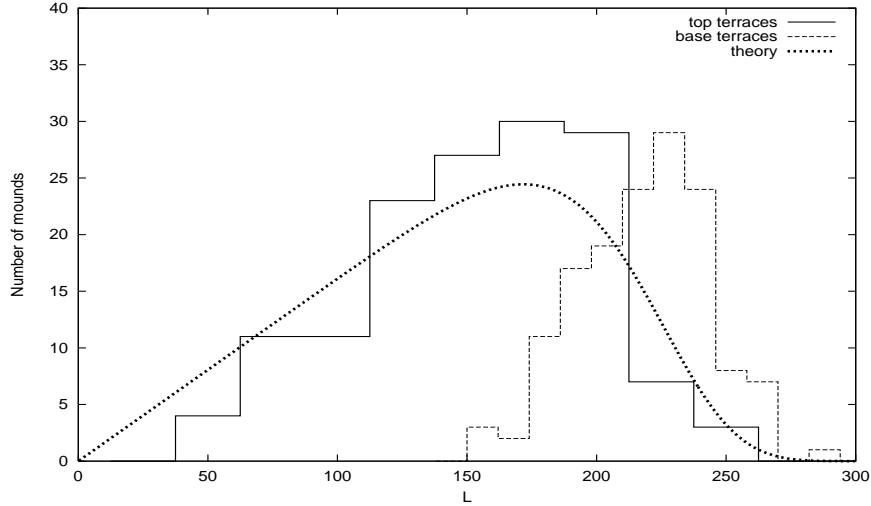,height=11.5cm,width=7.cm,angle=-90}}
\caption{Distribution of sizes of top terraces, measured in terms of 
their perimeter $L$ in units of lattice sites \protect\cite{Krug00a}.
The full line shows a histogram obtained from the evaluation of STM images
of 145 mounds, grown at 440 K in the presence of a partial CO pressure
of $1.9 \times 10^{-9}$ mbar, at a total coverage of 37.1 monolayers.
The thick dotted line shows the (appropriately normalized) 
prediction (\ref{P(R)}). 
The thin dashed line shows the corresponding histogram for the
base terraces, i.e. the terraces supporting the top terraces. 
It has been included to demonstrate that the assumption of 
a constant size of base terraces is reasonable.}
\end{figure}

Within this model,
the probability $P(R)$ 
to observe a top terrace of size $R$ is equal to 
the probability that the time since the last nucleation event 
is \emph{at least} $F^{-1} (R/R_b)^2$, which equals
$P_0(F^{-1} (R/R_b)^2)$. Changing variables from $t$ to $R$ then yields
\begin{equation}
\label{P(R)}
P(R) = \frac{2 R}{F R_b^2} \exp \left(- 
\frac{\pi^2 F R^7}{7 \nu' R_b^2} \right).
\end{equation}
As can be seen in Fig.10, this gives a quite
satisfactory description
of the experimental data. By fitting the mean of (\ref{P(R)}), 
the step edge barrier for Pt(111) in the presence of CO 
has been extracted from images
of multilayer mounds \cite{Krug00a}. 
The result $\Delta E_{\mathrm S} \approx 0.33$ eV is in good agreement
with the estimate of 0.36 eV obtained from second layer nucleation in the
submonolayer regime (see Section \ref{2nd}). 

\subsection{THE TOP TERRACE SIZE FOR REVERSIBLE AGGREGATION}

The derivation of (\ref{Rc}) can be repeated using the expressions 
(\ref{omega1},\ref{omega2}) for the nucleation
rate in the case of reversible aggregation, $i^\ast > 1$
\cite{Krug00b}.
In general the top terrace size is given by an expression of the
form 
\begin{equation}
\label{Rcgen}
R_c €\sim \left( \frac{\nu}{F} \right)^{\gamma'}
\left( \frac{\nu'}{\nu} \right)^{\mu'},
\end{equation}
where the exponents $\gamma'$ and $\mu'$ take the values
\begin{equation}
\label{Rc1}
\gamma' = \mu' = \frac{i^\ast}{3 i^\ast + 2}  
\;\;\; ({\mathrm regime \; II})
\end{equation}
\begin{equation}
\label{Rc2}
\gamma' = \frac{i^\ast}{i^\ast + 3}, \;\;\;
\mu' = \frac{i^\ast + 1}{i^\ast + 3} 
\;\;\; ({\mathrm regime \; III}) 
\end{equation}
In terms of the kinetic rates $\nu$, $\nu'$ and $F$, the two
regimes are defined by the inequalities 
$F/\nu \ll \nu'/\nu \ll (F/\nu)^{\delta_1}$ (regime II) and
$(F/\nu)^{\delta_1} \ll \nu'/\nu \ll (F/\nu)^{\chi/2}$
(regime III), 
where 
\begin{equation}
\label{delta1}
\delta_1 = \frac{i^\ast(2 i^\ast - 1)}{2 i^\ast (i^\ast + 1) + 2}
\end{equation}
and 
\begin{equation}
\chi = \frac{i^\ast}{i^\ast + 2}
\end{equation} 
is the island density
exponent of classical nucleation theory, which relates
the density $N$ of first layer islands to the ratio $F/\nu$ through 
\cite{Venables84}
\begin{equation}
\label{N}
N \sim \left( \frac{\nu}{F}\right)^\chi.
\end{equation}

\section{Continuum Theory}
\label{Continuum}

The continuum theory of mound formation is based on the notion 
of a growth-induced, uphill surface current 
\cite{Villain91,Krug97b,Krug93,Politi00}. Figure 11
illustrates the basic idea: Because atoms deposited onto 
a vicinal terrace on the side of a mound attach preferentially
to the ascending step, they travel on average in the direction of the
uphill slope\footnote{It is important to realize that this does 
\emph{not} require atoms to actually climb uphill across steps
\cite{Krug93}.}
between the point of deposition and the point of incorporation.

\begin{figure}
\label{Current}
\centerline{\epsfig{figure=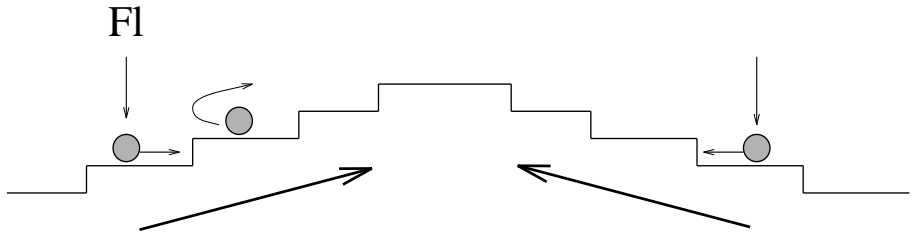,height=3.cm,width=12.cm,angle=0}}
\caption{Illustration of the uphill surface current generated
by step edge barriers.}
\end{figure}

The evaluation of the current is particularly simple when 
interlayer transport is completely suppressed, 
and nucleation on the vicinal terrace can be 
neglected\footnote{Expressions for more general situations are
derived in \cite{Krug97b,Politi00}.}. An atom deposited
onto a vicinal terrace of width $l$ then travels an average
distance $l/2$ to the ascending step, and hence the current
is $F l/2$. Describing the surface profile by a continuous height 
function  $h({\vec r},t)$ which measures the film thickness
(in units of the monolayer thickness) above a substrate point 
${\vec r}$, the local terrace width is $l = \vert \nabla h 
\vert^{-1}$, and hence the current is given by the expression
\begin{equation}
\label{J}
\vec J = \frac{F}{ 2 \vert \nabla h \vert^2} \nabla h.
\end{equation}
The surface profile evolves according to the continuity
equation
\begin{equation}
\label{cont}
\frac{\partial h}{\partial t} + \nabla \cdot \vec J = F.
\end{equation}
Here we are specifically interested in radially symmetric mounds.
Rewriting (\ref{cont}) in polar coordinates and using (\ref{J})
yields the following evolution equation for the mound profile
$h(r,t)$:
\begin{equation}
\label{hr}
\frac{\partial h}{\partial t} = - \frac{F}{2r}  
\frac{\partial}{\partial r} \; r \left(\frac{\partial h}{\partial r}
\right)^{-1} + F.
\end{equation}
We want to show that (\ref{hr}) possesses solutions corresponding
to the asymptotic mound shape derived in Section \ref{Asympt}.
To this end we make the ansatz
\begin{equation}
\label{hscale}
h(r,t) = \sqrt{Ft} \; \psi(r) + Ft.
\end{equation}
Inserting this into (\ref{hr}) yields the differential equation
\begin{equation}
\label{psi}
\frac{d}{dr} \left( \frac{r}{\psi'(r)} \right)
= r \psi(r).
\end{equation}
In order to establish the equivalence between the shapes
described by $\psi(r)$ and the scaled coverage distribution
$\Phi(x)$ of Section \ref{Asympt}, we need to verify that
the function $\psi(r)$ defined implicitly by
\begin{equation}
\label{equiv}
\left( \frac{r}{R_0} \right)^2 = \Phi(\psi(r)) 
\end{equation}
solves (\ref{psi}). Indeed, taking the derivative with 
respect to $\psi$ on both sides yields
\begin{equation}
\label{equiv2}
r \left( \frac{d\psi}{dr} \right)^{-1} = 
- \frac{C R_0^2}{2} e^{-\psi^2/2}
\end{equation} 
which reduces to (\ref{psi}) upon taking another derivative
with respect to $r$.

This calculation shows explicitly that the sides of the mounds
evolve according to the continuum equation (\ref{cont}); 
in \cite{Krug97} the same result was obtained for a one-dimensional
geometry. The continuum description does however not include
the nucleation on the top terraces, which have to be added to 
the profile determined by (\ref{psi}) as a boundary condition.
The development of a continuum theory of epitaxial growth which
explicitly incorporates nucleation remains a challenge for the
future.

\vspace{0.5cm}

\noindent
{\bf Acknowledgements.} Part of this paper is based on joint
work with Thomas Michely and Paolo Politi.
Support by DFG within
SFB 237 \emph{Unordnung und grosse Fluktuationen}
is gratefully acknowledged.

\end{document}